\documentclass[conference]{IEEEtran}
\IEEEoverridecommandlockouts
\usepackage[T1]{fontenc}
\usepackage{graphicx}
\usepackage[nobreak]{cite}
\usepackage{url}
\usepackage{xspace}
\usepackage{xcolor}
\usepackage{multirow}
\usepackage{amsmath}
\usepackage{amssymb}
\usepackage{newtxmath}

\usepackage{cleveref}
\crefname{figure}{Fig.}{Figs.}
\Crefname{figure}{Figure}{Figures}
\crefname{table}{Table}{Tables}
\crefname{section}{Section}{Sections}

\makeatletter
\let\MYcaption\@makecaption
\makeatother
\usepackage[font=footnotesize]{subcaption}
\makeatletter
\let\@makecaption\MYcaption
\makeatother

\newcommand{\TP}{\mathit{TP}}
\newcommand{\TN}{\mathit{TN}}
\newcommand{\FP}{\mathit{FP}}
\newcommand{\FN}{\mathit{FN}}
\newcommand{\cm}{\mathit{cm}}
\newcommand{\bG}[1]{{\overline{G_{#1}}}}
\newcommand{\bP}[1]{{\overline{P_{#1}}}}
\def\Methods{\mathit{methods}}

\newcommand{\V}[1]{\phantom{#1}}
\newcommand{\Proj}[1]{\texttt{#1}}

\newcommand{\RQ}[1]{\textit{RQ}${}_{\mathrm{#1}}$}
\newcommand{\rqOne}{How does performance vary between class-level prediction and method-level prediction when evaluated at their respective levels?}
\newcommand{\rqTwo}{How does performance vary between class-level prediction and method-level prediction when evaluated at method-level?}
\newcommand{\rqThree}{How does performance vary between class-level prediction and method-level prediction in effort-aware evaluation?}

\usepackage{framed}
\newcommand{\Conclusion}[1]{\begin{framed}\noindent #1\end{framed}}

\begin{document}

\title{
Revisiting Method-Level Change Prediction: A Comparative Evaluation at Different Granularities
}

\author{\IEEEauthorblockN{Hiroto Sugimori}
\IEEEauthorblockA{%
  \textit{School of Computing}\\\textit{Institute of Science Tokyo}\\
  Meguro-ku, Tokyo 152--8550, Japan\\
  sugimori@se.comp.isct.ac.jp}
\and
\IEEEauthorblockN{Shinpei Hayashi}
\IEEEauthorblockA{%
  \textit{School of Computing}\\\textit{Institute of Science Tokyo}\\
  Meguro-ku, Tokyo 152--8550, Japan\\
  hayashi@comp.isct.ac.jp}
}

\maketitle
\pagestyle{plain}
\thispagestyle{plain}

\begin{abstract}
To improve the efficiency of software maintenance, change prediction techniques have been proposed to predict frequently changing modules.
Whereas existing techniques focus primarily on class-level prediction, method-level prediction allows for more direct identification of change locations.
Method-level prediction can be useful, but it may also negatively affect prediction performance, leading to a trade-off.
This makes it unclear which level of granularity users should select for their predictions.
In this paper, we evaluated the performance of method-level change prediction compared with that of class-level prediction from three perspectives: direct comparison, method-level comparison, and maintenance effort-aware comparison.
The results from 15 open source projects show that, although method-level prediction exhibited lower performance than class-level prediction in the direct comparison, method-level prediction outperformed class-level prediction when both were evaluated at method-level, leading to a median difference of 0.26 in accuracy.
Furthermore, effort-aware comparison shows that method-level prediction performed significantly better when the acceptable maintenance effort is little.
\end{abstract}
  
\section{Introduction}\label{s:introduction}
Software continuously evolves even after its release.
The reasons for changes include changing requirements, refactoring, and bug fixes\cite{lehman-change}.
These changes are essential for improving the software quality.
However, numerous changes can lead to increased software complexity.
Therefore, developers need to implement changes while preventing software complexity.

Studies on software change prediction are being conducted to support software change activities\cite{malhotra-survey,catolino-pre,silva-history,catolino-ensemble,elish-process,catolino-dev,Lu-oo,farah-method-level,catolino-smell,mojeeb-network}.
The goal of this task is to improve the software quality by predicting change-prone modules that are prone to frequent changes in the next release.
By identifying these change-prone modules, developers can refactor the relevant source code and allocate human resources according to the anticipated amount of change. 
Therefore, developers can implement changes smoothly while preventing the software from becoming more complex.
In addition, the developer's task context\cite{mylyn} can be inferred based on the change prediction results.
To reduce information overload in development, a development environment\footnote{https://eclipse.dev/mylyn} has been proposed to display only the modules that are relevant to the tasks on which the developer is working, based on the development context\cite{mylyn,mylar}.
If change-prone modules can be accurately identified, developers can concentrate only on the modules relevant to their tasks.
Additionally, based on this inferred context, they can also plan refactoring that will contribute to future code changes in development tasks\cite{natthawute-mtd2017,natthawute-jsep201806}.

Existing change prediction approaches focus primarily on the class-level prediction\cite{catolino-pre,catolino-ensemble,silva-history,elish-process,catolino-dev,Lu-oo,catolino-smell,mojeeb-network}.
However, the size of a class can range from hundreds to thousands of lines. 
Therefore, it is unclear where exactly changes will occur within a class that is predicted to be change-prone.

Method-level prediction, which offers a finer granularity, enables developers to identify where changes will occur more directly\cite{farah-method-level}. 
The usefulness of method-level prediction can be clarified when a developer performs method-level refactoring to facilitate future changes based on class-level predictions.
The developer refactors several methods that belong to a class predicted to be change-prone.
If changes occur in other methods within that class that were not refactored, the refactoring effort of the developer will prove ineffective.
Accurate method-level prediction, i.e., yielding results with fewer noisy methods, allows us to refactor the methods where changes actually occur.
Method-level prediction enables a more direct identification of change locations; however, it may also lead to a reduced prediction performance.
A trade-off may exist between the granularity of predictions and their performance, leaving developers uncertain about which level to choose in practice.

This paper aims to clarify the effectiveness of method-level change prediction by comparing the results at class-level and method-level.
We designed our prediction technique based on the method-level prediction study by Farah et al.\cite{farah-method-level}.
To investigate the usefulness of method-level change prediction, we derived the following research questions (RQs):
\begin{itemize}
  \item \textbf{\RQ{1}}: \emph{\rqOne}
  \item \textbf{\RQ{2}}: \emph{\rqTwo} 
  \item \textbf{\RQ{3}}: \emph{\rqThree} 
\end{itemize}
The motivation for each RQ is explained in detail later.
By answering these RQs, we clarify the usefulness of method-level change prediction.

The main contributions of this paper are shown below.
\begin{itemize}
    \item We reproduced the method-level change prediction technique proposed by Farah et al.
    \item We re-evaluated method-level change prediction by comparing it with class-level prediction from three perspectives: direct, method-level, and effort-aware comparisons.
    \item We provided guidelines for the context in which method-level prediction methods should be used based on the comparison results. 
\end{itemize}

The investigation showed that when class-level and method-level predictions were directly evaluated at their respective levels, method-level prediction showed lower accuracy.
However, when both were evaluated at method-level, method-level prediction outperformed class-level prediction across multiple evaluation metrics.
In addition, effort-aware comparison showed that method-level prediction performed significantly better when the acceptable maintenance effort was little.

The remainder of this paper is organized as follows.
\Cref{s:rwork} provides an overview of existing studies related to this work.
\Cref{s:technique} presents the details of our change prediction technique.
\Cref{s:empirical} reports the study design and results of the empirical studies, and \cref{s:discussion} provides insights from the experiments.
\Cref{s:validity} examine the threats to the validity of this paper.
Finally, we conclude this study and state our future work in \cref{s:conclusion}.

\section{Related Work}\label{s:rwork}
\begin{table}[tb]
    \caption{Comparison of Related Work}\label{tab:rwork}
    \centering
        \begin{tabular}{llll} \hline
              & Class-level & Method-level  & Comparison\\ \hline
            Bug prediction  & $\checkmark$ \cite{ckbug,designbug} & $\checkmark$\cite{Hata-method-bug,Giger-method-bug} & $\checkmark$\cite{Hata-method-bug}\\
            Bug localization & $\checkmark$ \cite{Almhana-class-bl,zhou-filebl} & $\checkmark$\cite{ZHANG-method-bl,tsumita-saner2023,YOUM-method-bl,amalgam} & $\checkmark$\cite{tsumita-saner2023} \\
            Change prediction & $\checkmark$ \cite{malhotra-survey,catolino-pre,catolino-ensemble,silva-history,elish-process,catolino-dev,Lu-oo,farah-method-level,catolino-smell,mojeeb-network} & $\checkmark$ \cite{farah-method-level} & No study.\\
            \hline
        \end{tabular}
\end{table}
Various change prediction techniques have been proposed to date\cite{malhotra-survey,catolino-pre,catolino-ensemble,silva-history,elish-process,catolino-dev,Lu-oo,farah-method-level,catolino-smell,mojeeb-network}. 
Most of these techniques employ machine learning models and are formulated as a classification problem.
They utilize product and process metrics as independent variables, with a binary label that indicates whether a module is change-prone, serving as the dependent variable.

In the related tasks, such as bug prediction and bug localization, researchers have developed class-level and method-level prediction techniques\cite{Almhana-class-bl, amalgam,ckbug,designbug,Giger-method-bug,Hata-method-bug,tsumita-saner2023,YOUM-method-bl,ZHANG-method-bl,zhou-filebl} and demonstrated effectiveness of method-level predictions by comparing them with class-level predictions\cite{Hata-method-bug,tsumita-saner2023}.
\cref{tab:rwork} presents examples of studies that have proposed techniques at each level, as well as studies that compare both levels.
Researchers have used the effort-aware evaluation to compare the two levels.
Although method-level prediction techniques have also been proposed for change prediction, comparisons with class-level predictions have yet to be conducted.

\subsection{Class-Level Change Prediction}

Lu et al. demonstrated the effectiveness of product metrics in change prediction through an empirical study involving 102 projects \cite{Lu-oo}.
Elish et al. derived process metrics to be used as inputs for change prediction using the Goal-Question-Metric approach and demonstrated their effectiveness through empirical experiments\cite{elish-process}. 
Furthermore, they revealed that predictions using both product and process metrics showed higher performance than those using only one type of metric.
Catolino et al. proposed a prediction technique that incorporates developer-related information, which is commonly used in bug prediction\cite{catolino-dev}.
They constructed models using product metrics, process metrics, and developer information as independent variables and compared the outcomes.
The model incorporating developer information performed better than the other models.
They also found that these metrics complement one another in terms of prediction performance.
In addition, they demonstrated that utilizing ensemble methods in change prediction markedly improves the prediction performance \cite{catolino-ensemble}.
Furthermore, they added code smell-related information as the independent variable to improve change prediction performance\cite{catolino-smell}.

\subsection{Method-Level Bug Prediction}
In a similar task of bug prediction, several method-level prediction techniques have already been proposed, and their usefulness has been demonstrated.
Giger et al. proposed a method-level bug prediction technique using product metrics and process metrics \cite{Giger-method-bug}. 
They employed product metrics that are definable at method-level in CK \cite{aniche-ck}, a product metric computation tool. 
In addition, they used ChangeDistiller \cite{ChangeDistiller} to gather method-level change histories to calculate the process metrics.
Hata et al. proposed a method-level bug prediction technique that uses process metrics \cite{Hata-method-bug}. 
They utilized a fine-grained repository \cite{Historage} to calculate method-level process metrics.
A fine-grained repository generated by their technique \cite{Historage} tracks software change histories with finer granularity.
They employed an evaluation approach that accounts for an effort to compare class-level and method-level predictions.
The findings indicated that method-level prediction is more efficient than class-level prediction in identifying bugs. 

\begin{table*}[tb]\centering
    \caption{Independent Variables}\label{tab:metrics}
    \begin{tabular}{llll} \hline
         & Granularity & Product Metrics & Process Metrics \\ \hline
        Farah et al. & Method-Level & CK\cite{aniche-ck} and Understand: 83 types & Elish et al.\cite{elish-process}: 17 types \\
        \hline
        \multirow{2}{*}{Our technique} & Class-Level & CK\cite{aniche-ck}: 49 types & Elish et al.\cite{elish-process}: 17 types  \\
        & Method-Level & CK\cite{aniche-ck}: 31 types & Elish et al.\cite{elish-process}: 17 types \\
        \hline
    \end{tabular}
\end{table*}

\subsection{Method-Level Change Prediction}
Similar to bug prediction, a method-level prediction technique has been proposed for change prediction. 
Farah et al. extended class-level change prediction technique using product and process metrics to method-level \cite{farah-method-level}.
Like Giger et al., they used ChangeDistiller \cite{ChangeDistiller} to collect method-level process metrics. 
In addition, they analyzed the metrics and machine learning models that are effective for change prediction.
The results of their empirical experiments demonstrated that prediction models, which incorporate both types of metrics, perform better than those that rely on a single type of metric.
To the best of our knowledge, this is the only paper that proposes a method-level change prediction technique.

In the above studies, class-level and method-level change prediction have not been compared, and the effectiveness of method-level change prediction has not been sufficiently demonstrated. 
In this paper, we replicated the technique of Farah et al. as closely as possible and conducted prediction experiments at both class-level and method-level.
We evaluated the usefulness of method-level change prediction by comparing the results.

\section{Technique}\label{s:technique}
In this paper, we conduct experiments at both class-level and method-level based on the change prediction technique proposed by Farah et al.\cite{farah-method-level}.
This technique was selected for two reasons.
First, to the best of our knowledge, their technique is the only one that addresses method-level change prediction.
Second, their machine learning approach, which uses product and process metrics, aligns with the basic framework of existing class-level prediction techniques\cite{malhotra-survey}.

\begin{figure}[tb]
  \centering
  \includegraphics[width=8cm]{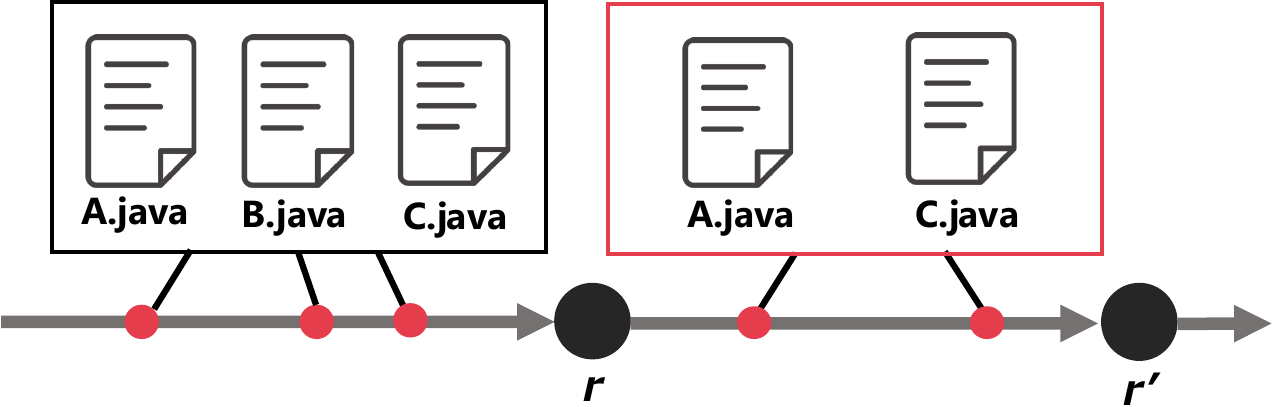}
  \caption{Changes between releases.} \label{fig:release}
\end{figure}
The software release scenario assumed in this paper is illustrated in \cref{fig:release}. 
In this scenario, we consider a project with multiple releases, where several changes occur in each release.
Let $r$ represent the release and $r'$ be the next release of $r$.
Our technique uses metrics that can be collected from the information prior to $r$ to predict the modules that will change between $r$ and $r'$.

\begin{figure*}[tb] 
  \centering
  \includegraphics[width=12cm]{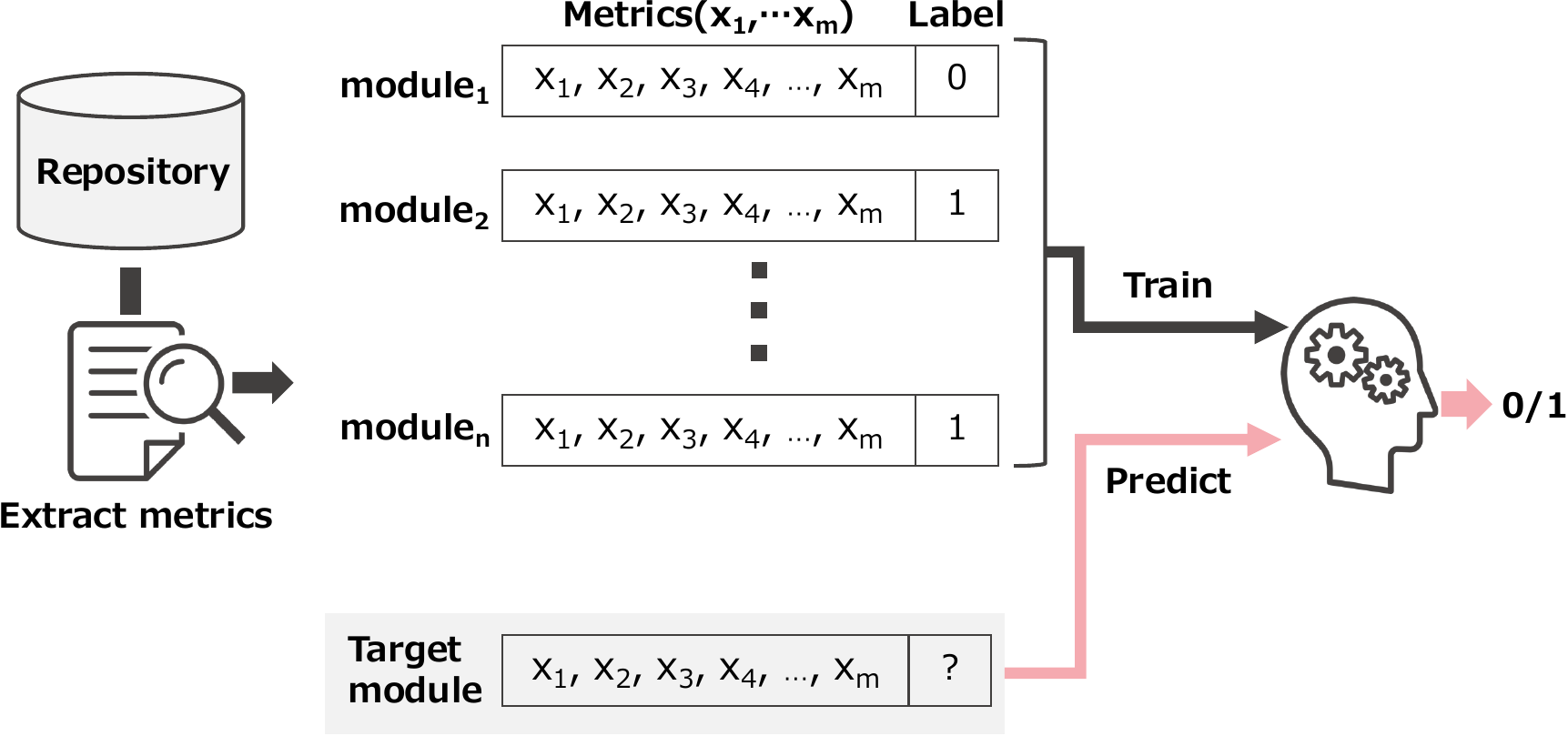} 
  \caption{Overview of prediction technique.} \label{fig:tech}
\end{figure*}

An overview of the prediction technique is shown in \cref{fig:tech}.
The repository of the target project is analyzed to calculate metric values for each module.
A dataset containing metric values \(x_1, x_2, \dots, x_m\) for the project modules serves as the independent variables alongside a binary (0/1) label that identifies whether or not a module is change-prone, which acts as the dependent variable.
This setup is used to learn the relationship between the independent and dependent variables.
When the metric values of a specific target module are input into the trained model, it predicts whether the module is likely to be change-prone.

In the remainder of this section, we first describe the technique proposed by Farah et al. and then explain our technique and how it differs from theirs.

\subsection{Technique by Farah et al.}
Farah et al. investigated the most effective techniques for method-level change prediction by applying various machine learning models and data resampling techniques and comparing their performance.
In this subsection, we briefly present the techniques they compared and their results.

\subsubsection{Independent Variables}
Farah et al. selected product and process metrics as independent variables for prediction based on their literature review.
They compared the results of using each metric individually with those of combining both metrics.
The number of types of metrics they used is listed in \cref{tab:metrics}.
The comparison showed that the use of both product and process metrics led to the highest prediction performance.
They used CK\cite{aniche-ck} and Understand\footnote{\url{https://scitools.com/}} to extract product metrics.
CK and Understand are static analysis tools used for calculating product metrics.
For extracting process metrics, they used a tool they developed themselves.

\subsubsection{Dependent Variable}
They defined modules with a number of changes greater than the median number of changes across all methods between releases as change-prone.
This definition was provided by Romano et al.\cite{daniele-label}.
They used ChangeDistiller\cite{ChangeDistiller} to extract method-level changes between releases.
This tool implements a diff algorithm that generates and compares syntax trees between two versions of a project.

\subsubsection{Data Preprocessing}
Before creating the machine learning model, they performed data preprocessing.
The values of the metrics in a numeric format were normalized between 0 and 1 using the min-max technique.
They found imbalances in the distribution of dependent variables in the data.
As data imbalances can adversely affect learning, they utilized the following six data resampling techniques to mitigate this issue: Synthetic Minority Oversampling Technique (SMOTE)\cite{smote}, Random Over-Sampler (ROS), Adaptive Synthetic (ADA), Random Under-Sampling (RUS), Edited Nearest Neighbors (ENN), and TomekLinks (TL).
As a result of comparing their performance, ROS and RUS exhibited the best performance.
They used scikit-learn\cite{sklearn} to implement these resampling techniques.

\subsubsection{Building Classifier}
They compared the results using Decision Tree, Logistic Regression, Multi-Layer Perceptron, and Random Forest.
These models have been widely used in existing research\cite{malhotra-survey}.
Among the aforementioned data resampling techniques, the best result was achieved when combined with Random Forest and RUS.

\subsection{Our Technique}\label{s:ourTech}
Based on the technique proposed by Farah et al., we designed prediction techniques at both the method-level and class-level.
Although they have published part of the program necessary for the experiment, the artifact was not designed to execute the whole process.
Therefore, we re-implemented the entire program to conduct experiments.
The purpose of this paper is not to reproduce their technique perfectly but to compare results between different levels.
Therefore, unlike their approach, we do not compare various prediction techniques but instead adopt the technique that demonstrated the best performance in their experiments. 

\subsubsection{Independent Variables}
We employed product and process metrics as independent variables because
Farah et al. demonstrated that using both metrics together yields better results than using either one alone.
The number of types of metrics we used for class-level prediction and method-level prediction is also shown in \cref{tab:metrics}.

Product metrics are calculated based on the source code at release $r$.
We used CK\cite{aniche-ck} to collect product metrics at both class-level and method-level.
Although Farah et al. computed more metrics with Understand, our approach relies solely on CK because of its open-source nature and lower implementation cost.
Since Understand is commercial software, it is not freely accessible to everyone.

Process metrics are calculated based on the commit history prior to $r$.
We used the same types of metrics set as Farah et al.
This set of process metrics was originally proposed by Elish et al. \cite{elish-process}.
We developed our own program to compute these metrics because the implementation used by Farah et al. for measuring the process metrics was not publicly available.
We used a method repository \cite{Historage} that allows simpler retrieval of method-level changes.
Method repository extracts code segments corresponding to the methods from each file in the repository and saves them as new files that carry over the method-level change history.
This allows the acquisition of method-level change histories, similar to how file-level change histories are captured using Git.
We used git-stein 
\cite{git-stein} to generate method repositories.

\subsubsection{Dependent Variable}\label{s:dependent}
As to Farah et al., modules with a number of changes greater than the median number of changes across all modules between releases are defined as change-prone.
While Farah et al. used ChangeDistiller to obtain method-level change histories, we did not adopt this approach in this paper.
Instead of ChangeDistiller, we used a method repository\cite{Historage} again here.
Differences in the tools utilized may lead to discrepancies in tracking renamed modules, although these are not substantial concerns in the context of machine learning-based techniques.

\subsubsection{Data Preprocessing}
Based on the research by Farah et al., we preprocessed the data prior to training a machine learning model.
The values of the metrics in a numeric format were normalized between 0 and 1 using the min-max technique.
We employed RUS as a data resampling technique.
According to Farah et al., RUS is the most effective data resampling technique when paired with the best machine learning model for change prediction, Random Forest.

\subsubsection{Classifier Building}
We employed Random Forest as our machine learning model because Farah et al. demonstrated that it is the most effective model for method-level prediction among traditional machine learning models.

\section{Empirical Study}\label{s:empirical}
We conducted experiments to answer the three research questions.

\subsection{Dataset Preparation}\label{s:dataset}
\begin{table*}[t]
    \caption{Dataset Overview}\label{tab:repo}
    \centering
        \begin{tabular}{llcccccc} \hline
             Original paper & Repositories & \# releases & \# commits & \# classes & \# methods & CP class & CP method\\ \hline

            \multirow{6}{*}{Farah et al.\cite{farah-method-level}}
            & \Proj{apache/commons-bcel} & 3 & 29--52 & 416--418 & 3,652--3,701 & 0.024--0.20\V{0} & 0.022--0.11\V{0}\\ 
            & \Proj{apache/pdfbox} & 10 & \V{0,0}43--6,180 & \V{0,}434--1,159 & \V{0}3,360--10,081 & 0.09--0.46 & 0.03--0.50\\
            & \Proj{apache/commons-io} & 15 & \V{0,0}34--1,022 & \V{0}39--242 & \V{0,}332--2,152 & 0.13--0.49 & 0.03--0.47 \\
            & \Proj{easymock/easymock} &  10 & \V{0}36--219 & \V{0}88--102 & 522--574 & 0.02--0.42 & 0.004--0.27\V{0}\\
            & \Proj{wro4j/wro4j} & 8 & \V{0,0}15--1,237 & 138--353 & \V{0,}685--1,847 & 0.03--0.47 & 0.06--0.49 \\
            & \Proj{igniterealtime/openfire} & 6 & 154--913 & 686--726 & 6,612--8,346 & 0.15--0.43 & 0.04--0.16\\
            \cline{2-8}
             & (Total) & 52 & \V{0,0}15--6,180 & \V{0,0}39--1,159 & \V{00,}332--10,081 & 0.02--0.49 & 0.03--0.50 \\
            \hline
                        \multirow{9}{*}{Catolino et al.\cite{catolino-smell}} 
            & \Proj{apache/logging-log4j1} & 2 & 226--365 & 86--87 & 455--518 & 0.43--0.45 & 0.41--0.42\\
            & \Proj{apache/ant} & 3 & 1,885--3,556 & 281--573 & 1,992--5,436 & 0.43--0.5\V{0} & 0.20--0.31 \\
            & \Proj{apache/ant-ivy} & 1 & 184 & 352 & 3,762 & 0.24 & 0.03 \\
            & \Proj{apache/camel} & 3 & \V{0,}498--1,142 & 153--447 & \V{0,}876--3,144 & 0.31--0.39 & 0.14--0.45 \\
            & \Proj{apache/forrest} & 1 & 874 & 29 & 165 & 0.38 & 0.17 \\
            & \Proj{apache/lucene} & 2 & 533--599 & 185--226 & 1,658--2,136 & 0.38--0.50 & 0.26--0.3\V{0}\\
            & \Proj{apache/poi} & 3 & 135--843 & 222--378 & 2,521--4,570 & 0.31--0.49 & \V{0}0.1--0.36\\
            & \Proj{apache/synapse} & 2 & 325--435 & 152--219 & 524--863 & 0.40--0.45 & 0.14--0.22\\ 
            & \Proj{apache/xerces2-j} & 2 & 239--453 & 262--172 & 2,164--2,289 & 0.36--0.48 & 0.12--0.14\\
            \cline{2-8}
             & (Total) & 19 & \V{0,0}84--3,556 & \V{0}29--573 & \V{0,}165--5,436 & 0.24--0.50 & 0.03--0.45\\
            \hline
            Total & & 71 & \V{0,0}15--6,180 & \V{0,0}29--1,159 & \V{00,}165--10,081 & 0.02--0.50 & 0.03--0.50\\
            \hline
        \end{tabular}
\end{table*}
In addition to the repositories selected by Farah et al., we used those selected by Catolino et al. \cite{catolino-smell} to confirm the applicability of our technique to other projects.
Catolino et al. have proposed a class-level change prediction technique using machine learning, similar to the technique proposed by Farah et al.
\begin{figure*}[tb] 
  \centering
  \includegraphics[width=17cm]{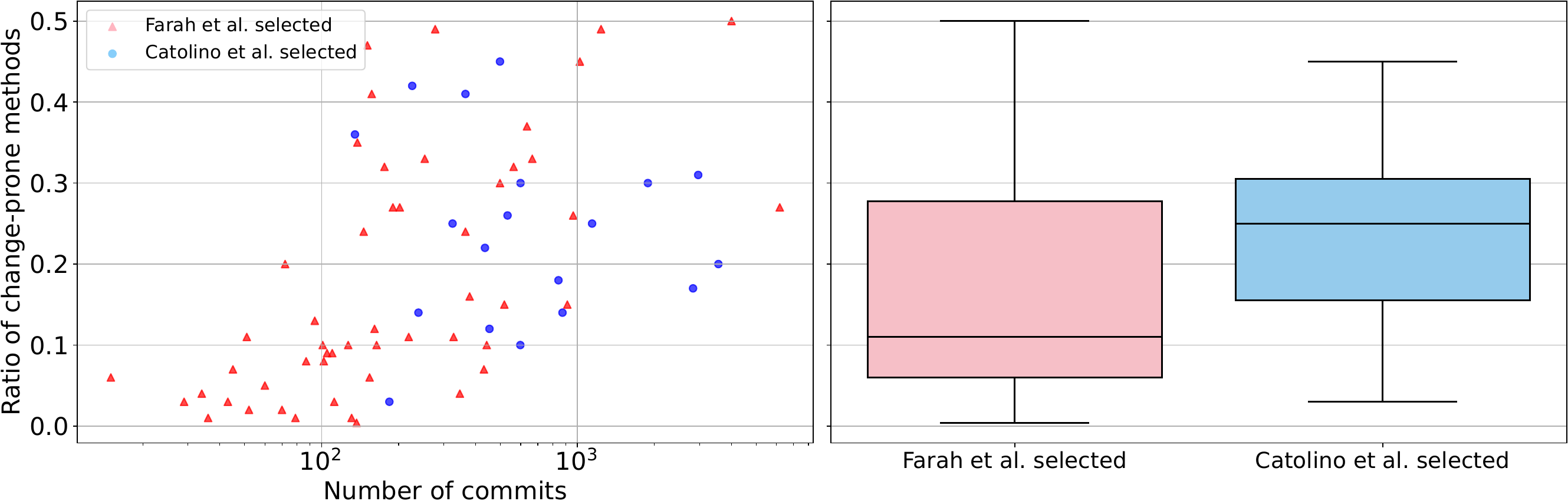} 
  \caption{Difference in the distribution of change-prone methods between the datasets by Farah et al. and Catolino et al.} \label{fig:dataset}
\end{figure*}

The repositories used and their main characteristics are shown in \cref{tab:repo}.
These repositories are open-source software projects developed in Java and hosted on GitHub.
In \cref{tab:repo}, the ``\# releases'' column represents the number of targeted releases in the repository. 
The ``\# commits'' column shows the range of commit counts for these releases. 
The ``\# classes'' column represents the range of a number of classes, and the ``\# methods'' column reflects the range of a number of methods. 
The ``CP class'' and ``CP method'' columns represent the ranges of the ratio of change-prone classes and change-prone methods, respectively.
Releases were identified using tags in the Git repository, and the commit history between tags was retrieved.
We found that the distribution trends of change-prone methods in the datasets used by Farah et al. and Catolino et al. differed.
We believe that this difference is owing to the number of commits, which represents the scale of the releases.
The scatter plot on the left in \cref{fig:dataset} shows the relationship between the proportion of change-prone methods and the number of commits for each release.
The box plot on the right shows the distribution of the proportion of change-prone methods for each release.
When the ratio of methods being changed is small, the median number of changes across all methods becomes zero.
In this case, as already mentioned in \cref{s:technique}, any method that is modified even once is considered change-prone; however, if this ratio is small, the percentage of change-prone methods is also low.

\subsection{Evaluation Metrics}\label{s:evaluation_metrics}
We employed evaluation metrics that are commonly used in change prediction, including the research by Farah et al.
First, we computed precision, recall, and accuracy defined as follows:
\begin{align*}
  \mathit{Precision} &= \frac{\TP}{\TP + \FP},\\
  \mathit{Recall} &= \frac{\TP}{\TP + \FN}, \\
  \mathit{Accuracy} &= \frac{\TP + \TN}{\TP + \TN + \FP + \FN}.
\end{align*}
where,  $\TP$ is the number of true positives,  $\TN$ is the number of true negatives, $\FP$ is the number of false positives, and $\FN$ is the number of false negatives.
F1-score is defined as the harmonic mean of precision and recall:
\begin{align*}
    \textrm{F1-score} = 2*\frac{\mathit{Precision}*\mathit{Recall}}{\mathit{Precision}+\mathit{Recall}}.
\end{align*}
Moreover, we employed the Area Under the ROC Curve (AUC-ROC).

\subsection{Replication}\label{s:replication}

\subsubsection{Motivation}
As explained in \cref{s:ourTech}, our technique is not a complete reproduction of that proposed by Farah et al.\cite{farah-method-level}. 
We conducted replication experiments to verify whether the prediction results from our technique did not differ substantially from those of Farah et al.

\subsubsection{Study Design}
We conducted experiments on method-level change prediction using the technique described in \cref{s:technique}.
The datasets were divided into those used by Farah et al. (upper half of \cref{tab:repo}) and those used by Catolino et al. (lower half of \cref{tab:repo}).
We then compared these results described in the paper by Farah et al.
This confirms that our technique does not substantially diverge from the approach by Farah et al. and can be replicated with other datasets.
For evaluation, we use the metrics explained in \cref{s:evaluation_metrics}, measured at method-level. 
However, since Farah et al. did not use precision for their evaluation, we calculated the values for the other evaluation metrics.

\subsubsection{Results and Discussion}
\begin{figure}[tb]\centering
    \begin{subfigure}{\linewidth}\centering
        \includegraphics[width=7cm]{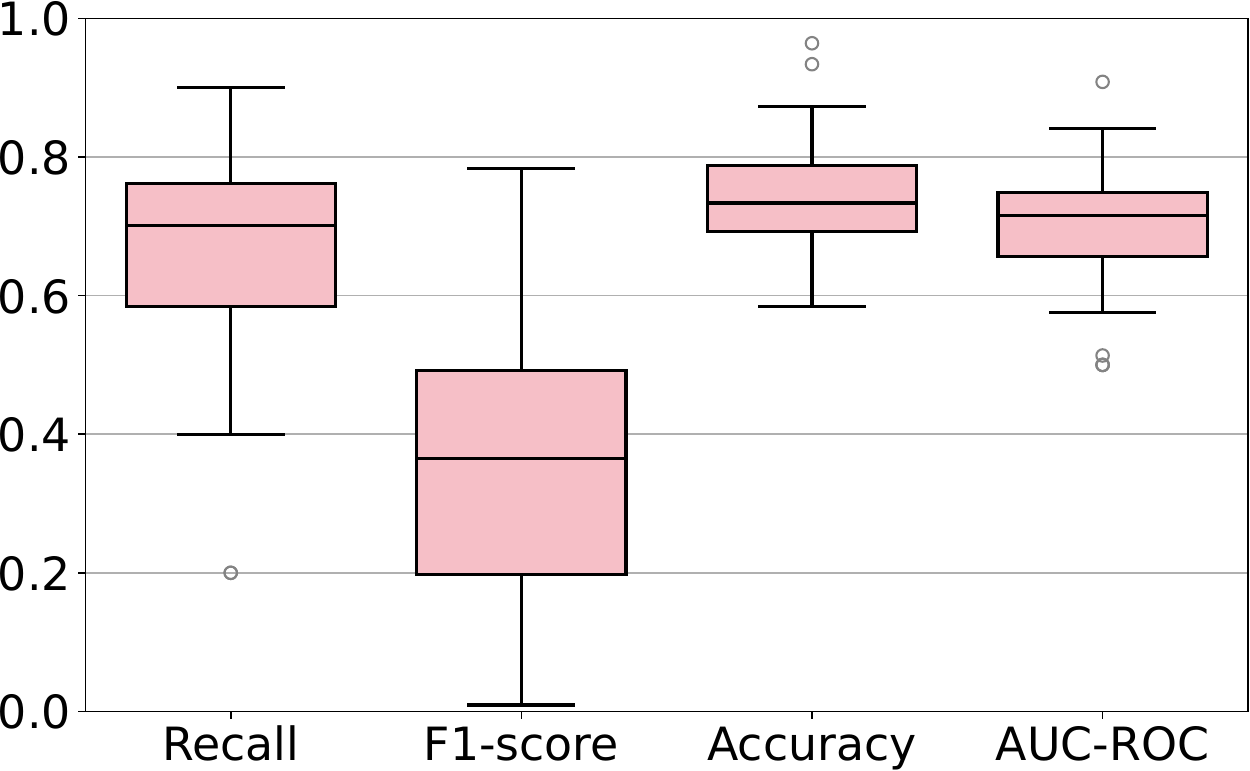}
        \caption{Replication on repositories used by Farah et al.}
        \label{fig:method-farah}
    \end{subfigure} ~ \\
    \begin{subfigure}{\linewidth}\centering
    \includegraphics[width=7cm]{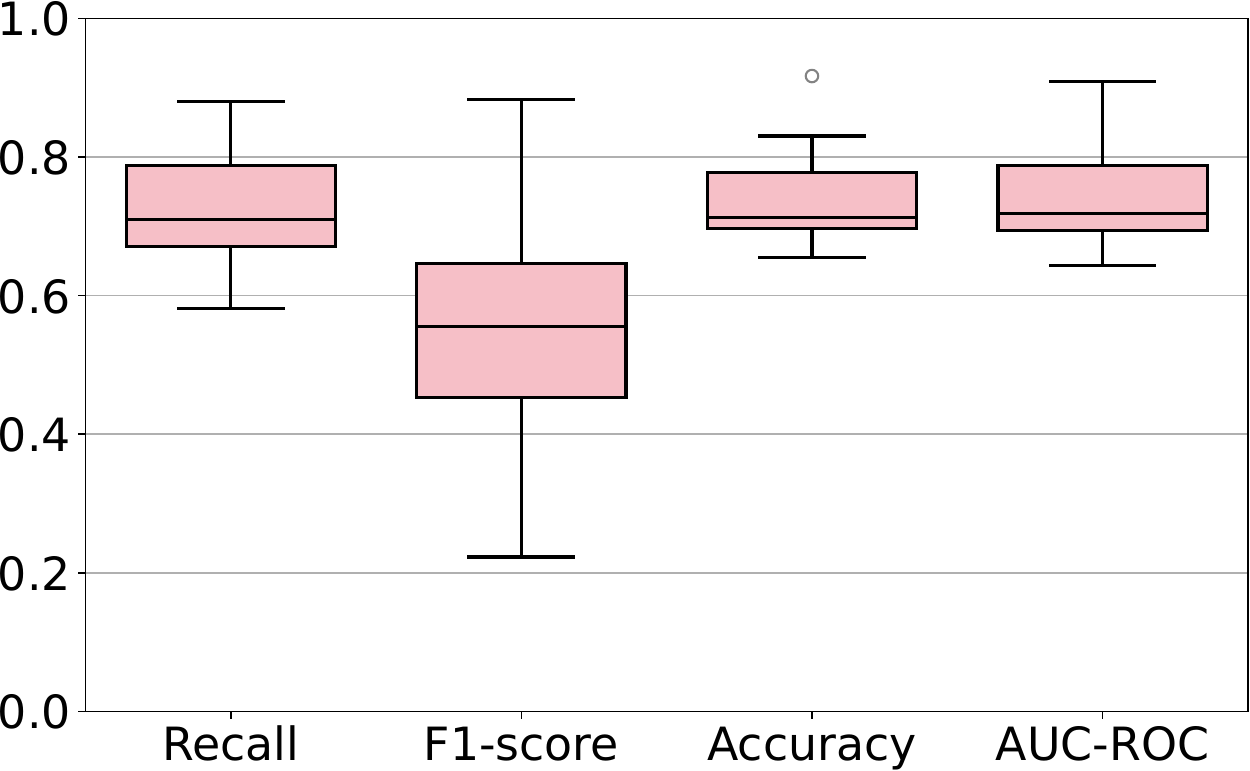}
    \caption{Replication on repositories used by Catolino et al.} \label{fig:method-catolino}
    \end{subfigure}
    \caption{Results of replication experiments.}
\end{figure}
The values of each evaluation metric from the experiments conducted on the dataset by Farah et al. are depicted in the box plots in \cref{fig:method-farah}.
These results were compared with those reported by Farah et al. 
They have published project-level results in their replication package rather than providing results for all releases of the target projects.
Therefore, we calculated our project-level results and compared them with theirs.
According to Farah et al., in comparing the performance of each model, we used the AUC-ROC as the main evaluation metric.
Among the repositories used by Farah et al., the largest AUC-ROC difference was 0.16 in the \Proj{igniterealtime/opefire} project, indicating that the values were close.
However, there was no substantial difference in the overall trend, and we believe that the variations are within the range attributable to the different sets of metrics used in our prediction compared with theirs.
The values of each evaluation metric in the experiments conducted on the dataset selected by Catolino et al. are also depicted in box plots in \cref{fig:method-catolino}.
Comparing these results with those of our experiments on the dataset selected by Farah et al., the trends in the metrics were similar; however, the F1-score increased slightly. 
As mentioned in \cref{s:dataset}, the dataset selected by Farah et al. has a smaller ratio of change-prone methods than that selected by Catolino et al., resulting in greater data imbalance.
We believe that this data imbalance degraded the performance of prediction on the dataset selected by Farah et al.
Based on these results, we believe that our technique is applicable across different datasets.

\Conclusion{Our technique differs from that proposed by Farah et al. in terms of the employed metrics, leading to different outcomes.
Among the repositories used by Farah et al., the largest AUC-ROC difference was 0.16.
Nevertheless, as the objective of our study is to compare different granularities, this issue does not detract from the subsequent discussion.}

\subsection{\RQ{1}: \rqOne}\label{s:rq1}
\subsubsection{Motivation}
Method-level prediction provides the location of changes more directly than class-level prediction.
However, there may be differences in performance due to the number of modules targeted for prediction and the metrics used for prediction varying between class-level and method-level predictions.
There could be a trade-off between the granularity of the predictions and their performance, leaving developers uncertain about which level to select in practice.
Therefore, we compare the performance of method-level prediction with class-level prediction.

\subsubsection{Study Design}
Following the technique described in \cref{s:technique}, we performed predictions at both class-level and method-level and compared their results.
As in \cref{s:replication}, the average value of each evaluation metric was calculated using 10-fold cross-validation.
We compared the results across the two levels using a two-sided Wilcoxon signed-rank test with a significance level of $\alpha = 0.05$.
We also reported Cliff’s delta ($d$) to measure the magnitude of the performance difference.
Cliff’s delta is interpreted based on the threshold proposed by Romano et al.\cite{cliff'sdelta}:
\emph{negligible} for $|d| < 0.147$,
\emph{small} for $0.147 \leq |d| < 0.33$,
\emph{medium} for $0.33 \leq |d| < 0.474$, and
\emph{large} for $0.474 \leq |d|$.

We provide the definitions of $\TP$, $\TN$, $\FP$, and $\FN$ for class-level and method-level.
Let the sets of all classes and methods in the target project be $E_c = \{\dots, e_c, \dots\}$ and $E_m = \{\dots, e_m, \dots\}$, respectively.
In class-level prediction, we define the ground truth of change-prone classes $G_c \subseteq E_c$ and $\bG{c} = E_c \setminus G_c$ as sets of classes that should be predicted to be change-prone and non-change-prone, respectively.
In the same way, in method-level prediction, we define the ground truth of change-prone methods $G_m \subseteq E_m$ and non-change-prone methods $\bG{m} = E_m \setminus G_m$.

Let $P_c \subseteq E_c$ and $P_m \subseteq E_m$ be the sets of the classes and methods predicted as change-prone, and $\bP{c} = E_c \setminus P_c$ and $\bP{m} = E_m \setminus P_m$ be those predicted as non-change-prone, respectively.
$\TP$, $\TN$, $\FP$, and $\FN$ at each level can be defined as follows:
\begin{align*}
  \TP_c &= |P_c \cap G_c|, &
  \TP_m &= |P_m \cap G_m|, \\
  \TN_c &= |\bP{c} \cap \bG{c}|, &
  \TN_m &= |\bP{m} \cap \bG{m}|, \\
  \FP_c &= |P_c \cap \bG{c}|, & 
  \FP_m &= |P_m \cap \bG{m}|, \\
  \FN_c &= |\bP{c} \cap G_c|, &
  \FN_m &= |\bP{m} \cap G_m|.
\end{align*}

\subsubsection{Results and Discussion}
\begin{figure}[tb] 
  \centering
  \includegraphics[width=8cm]{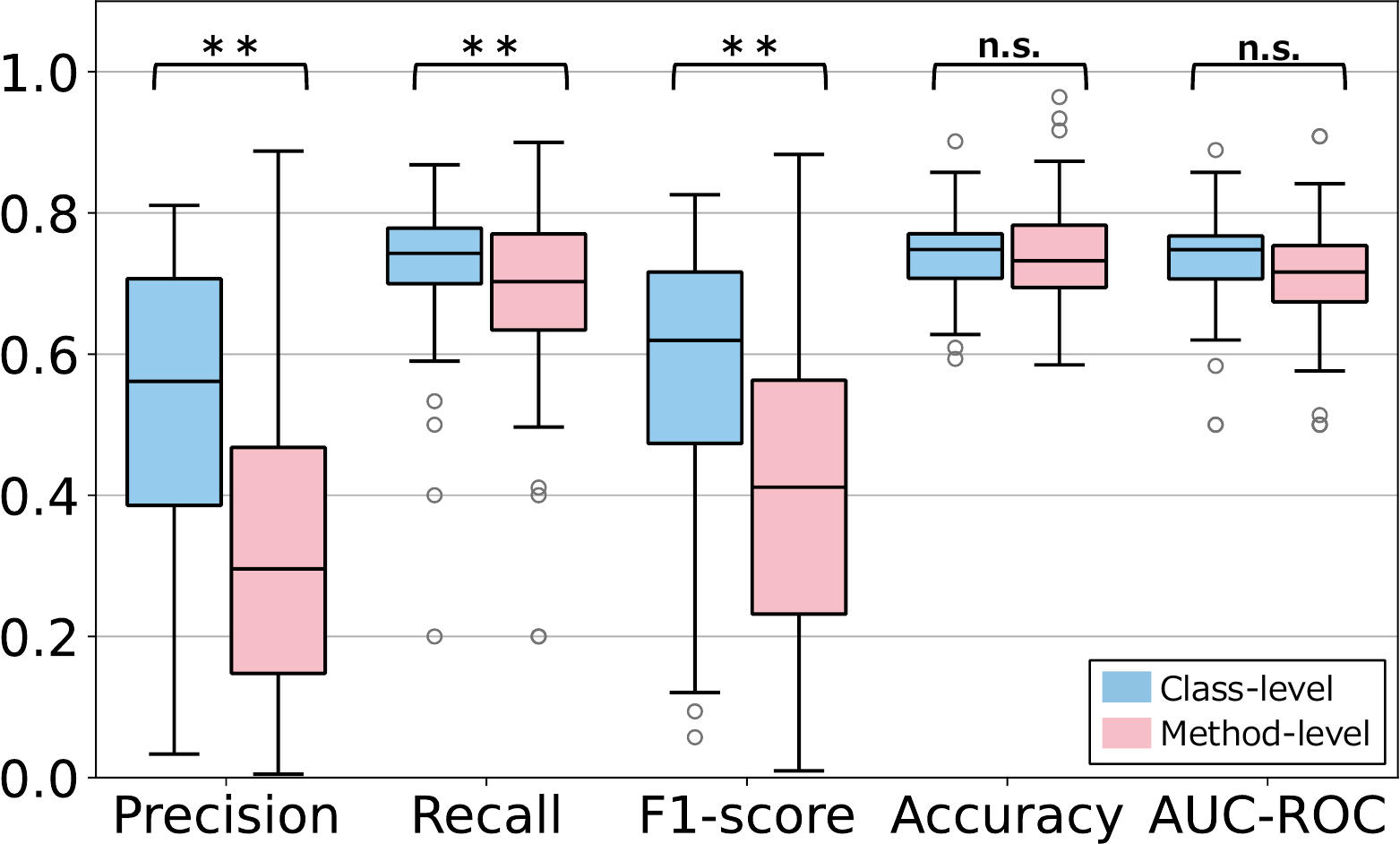} 
  \caption{Results for \RQ{1}: direct comparison.} \label{fig:results-rq1}
\end{figure}

\Cref{fig:results-rq1} shows the values for each evaluation metric at method-level and class-level, respectively.
\Cref{tab:rq1-result} presents the median value of each metric at both levels, the p-value from the Wilcoxon signed-rank test, as well as the values and results of Cliff’s delta.
In \cref{fig:results-rq1} and \cref{tab:rq1-result}, the statistical significance is denoted as follows: $p < 0.05$ is marked with $*$, $p < 0.01$ with $**$, and non-significant results are indicated by n.s.
This notation is used consistently in subsequent figures and tables throughout this paper.
The results suggested that class-level predictions outperformed method-level predictions in performance, and their differences were statistically significant for all metrics except for accuracy and AUC-ROC.
The median F1-score for class-level prediction was 0.61, while for method-level prediction, it was 0.2 points lower, at 0.41.
The effect sizes for precision and F1-score were large, while that for recall was small.

\begin{table}[tb]
    \caption{Results for \RQ{1}: Direct Comparison}\label{tab:rq1-result}
    \centering
        {\tabcolsep=2.5pt\begin{tabular}{lcccrl} \hline
             & Class-level & Method-level & p-value & \multicolumn{2}{c}{Cliff's delta} \\ \hline
            Precision & 0.56 & 0.29 & $1.0\times 10^{-13}$** & $-0.47$ & (large) \\ 
            Recall & 0.74 & 0.70 & $7.5\times10^{-4}$** & $-0.22$ & (small) \\
            F1-score & 0.61 & 0.41 & $5.2\times10^{-13}$** & $-0.48$ & (large)\\
            Accuracy & 0.74 & 0.73 & 0.24 n.s.& $-0.095$ & (negligible)\\
            AUC-ROC & 0.74 & 0.71 & 0.11 n.s.& $-0.20$ & (small)\\
            \hline
        \end{tabular}}
\end{table}

\Conclusion{Method-level prediction demonstrated lower performance compared to class-level prediction. 
In terms of the F1-score, the median of the method-level prediction result was 0.20 lower than that of the class-level prediction results, and its effect size was large.}

\subsection{\RQ{2}: \rqTwo}\label{s:rq2}

\subsubsection{Motivation}
In a change-prone class, there may be methods that are not change-prone (and vice versa).
The evaluation in answering \RQ{1} ignores such methods.
Therefore, we investigated the performance when considering class-level predictions at method-level.

\subsubsection{Study Design}
We re-evaluated the class-level predictions in \RQ{1} based on method-level ground truth and compared them with the method-level prediction results in \RQ{1}. 
In order to assess class-level predictions at method-level, we define $\TP$, $\TN$, $\FP$, and $\FN$ based on method-level ground truth.

We regard all the methods in all classes predicted as change-prone as change-prone methods: 
\[ P_\cm\,(\subseteq E_m) = \bigcup_{c \in P_c} \Methods(c) \]
where $\Methods(c)$ denotes the set of methods belonging to the class $c$.
Using the set of ground-truths, these numbers could be computed:
\begin{align*}
  \TP_\cm &= |P_\cm \cap G_m|, &
  \TN_\cm &= |\bP{\cm} \cap \bG{m}|, \\
  \FP_\cm &= |P_\cm \cap \bG{m}|, & 
  \FN_\cm &= |\bP{\cm} \cap G_m|.
\end{align*}
Here, $\bP{\cm} = E_m \setminus P_\cm$ represents the set of all methods included in the classes predicted as non-change-prone.

Using these values, we calculate precision, accuracy, recall, and F1-score for re-evaluating class-level predictions based on method-level ground truth.
To determine whether there was a difference in the median values of each evaluation metric between class-level predictions and method-level predictions, we applied the two-sided Wilcoxon signed-rank test at a significance level of 0.05.

\subsubsection{Results and Discussion}
\begin{figure}[tb]\centering

  \includegraphics[width=\linewidth]{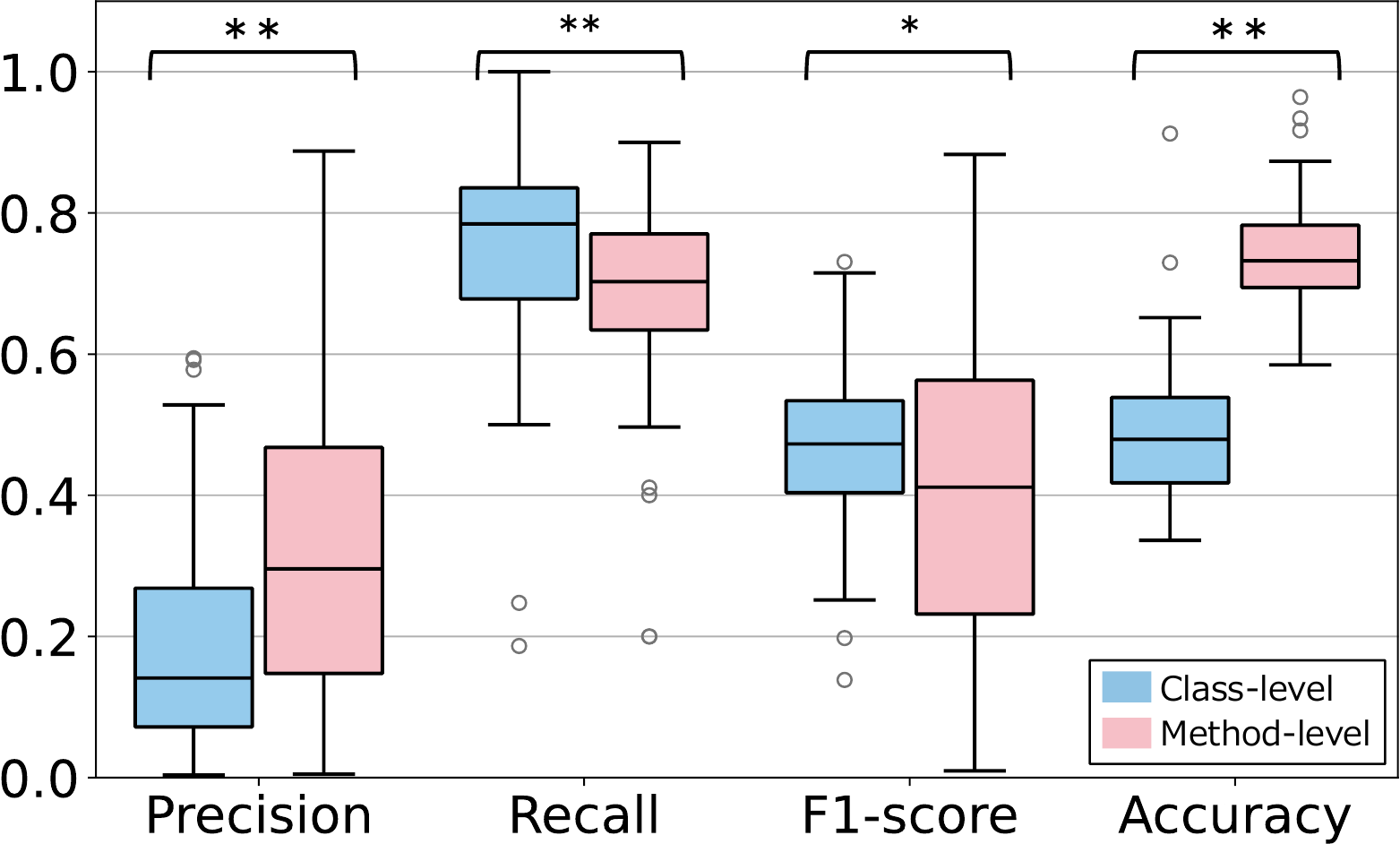} 
  \caption{Results for \RQ{2}: method-level evaluation.} \label{fig:method-class-to-method}
\end{figure}

\Cref{fig:method-class-to-method} shows the values of each evaluation metric at method-level and class-level, respectively.
\Cref{tab:rq2-result} shows the median values of each metric at each level, the p-values from the Wilcoxon signed-rank test, and the values and effect sizes of Cliff's delta.
These results exhibit a trend different from those of \RQ{1}.
Method-level predictions significantly outperformed in terms of precision and accuracy with medium and large effect sizes, respectively.
However, there was a decrease in the recall and F1-score with a small effect size.
These facts indicate that method-level predictions increase the number of false negative methods but reduce the number of false positive methods compared with class-level predictions.

\begin{table}[tb]\centering
    \caption{Results for \RQ{2}: Method-Level Evaluation}\label{tab:rq2-result}
    {\tabcolsep=3pt\begin{tabular}{lcccrl} \hline
         & Class-level & Method-level & p-value & \multicolumn{2}{c}{Cliff's delta} \\ \hline
        Precision & 0.14 & 0.29 & $8.3\times10^{-7}$**  & 0.36 & (medium) \\
        Recall & 0.78 & 0.70 & 0.0020** & $-0.33$ & (medium) \\
        F1-score & 0.47 & 0.41 & 0.013* & $-0.2$ & (small) \\
        Accuracy & 0.47 & 0.73 & $2.3\times10^{-17}$** & 0.94 & (large)  \\
        \hline
    \end{tabular}}
\end{table}

\Conclusion{Method-level prediction performed significantly better in terms of accuracy and precision, and their effect size was large and medium, respectively.  
The difference in median accuracy was 0.26. 
However, in terms of recall and F1-score, class-level prediction performed better.
These facts indicate that method-level prediction can reduce false positives; however, the false negatives are larger than those for class-level prediction.
}

\subsection{\RQ{3}: \rqThree}\label{s:rq3}

\subsubsection{Motivation}
Developers aim to utilize their limited efforts as efficiently as possible when maintaining software.
They often prioritize refactoring areas where future changes are anticipated over areas with severe code smells \cite{natthawute-jsep201806}.
An effort-aware evaluation was employed to compare the effectiveness of class-level prediction and method-level prediction in supporting efficient maintenance activities.

\subsubsection{Study Design}
We introduced an effort-aware evaluation to change prediction.
This evaluation technique allows for comparisons between different levels in terms of effort efficiency.
We adopt the approach of evaluating effort using Lines of Code (LOC), which is a common metric in the field of bug prediction\cite{Hata-method-bug}.
We assessed the effectiveness of class-level and method-level prediction using this approach when developers perform maintenance tasks on a given number of lines of source code based on change prediction results.
Unlike existing evaluation approaches that consider prediction results as a binary indicator of whether or not the modules are change-prone, we employed a probability-based ranking.
We then assess the usefulness of the ranking by considering how developers utilize the top-$k$ lines of modules in the ranking.
\begin{figure}[tb]\centering
  \begin{subfigure}{\linewidth}\centering
    \includegraphics[width=8cm]{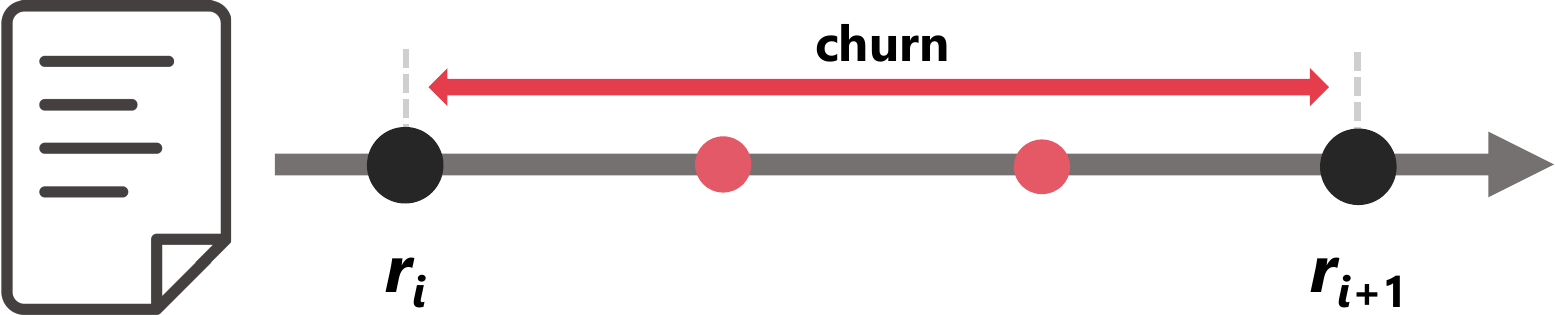} 
    \caption{Release-based change size.} \label{fig:release-based}
  \end{subfigure} ~ \\
  \begin{subfigure}{\linewidth}\centering
    \includegraphics[width=8cm]{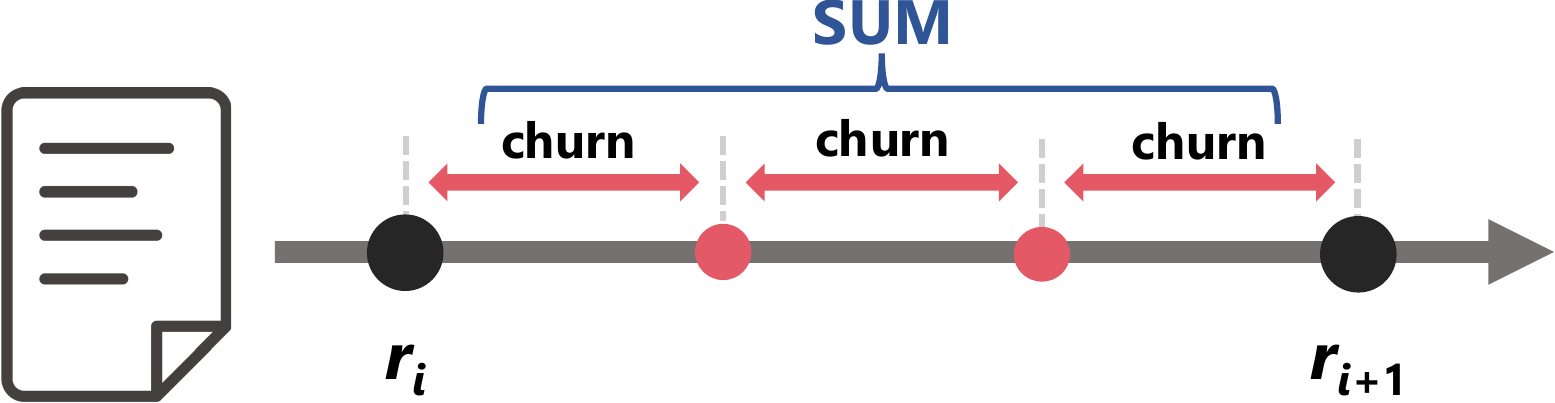} 
    \caption{Commit-based change size.} \label{fig:commit-based}
  \end{subfigure}
  \caption{Change size metrics.}\label{fig:change size metrics}
\end{figure}
Developers prefer to refactor code snippets that are likely to be changed in the future rather than those with severe code smells \cite{natthawute-jsep201806,natthawute-ieicet201807}.
Therefore, we argue that predictions for modules with more changes are useful.

For the purpose of quantifying change size, we define two types of metrics: \emph{release-based change size} and \emph{commit-based change size}, to measure the change size between releases from different perspectives.
\Cref{fig:change size metrics} illustrates their concepts.
A possible way is to compute the size of the differences between two snapshots of the releases.
\emph{Release-based change size} represents the sum of the \emph{code churn}~\cite{churn}, i.e., the total number of added and deleted lines, when comparing the beginning and ending snapshots of the release milestone for each module.
However, the changes implemented between releases may overlap, and the same code location may be modified multiple times. 
Therefore, the difference between the snapshots may underestimate the repeated efforts to the same code location. 
To mitigate this issue, we also computed the  \emph{commit-based change size}, which uses the sum of the churn of the changes of each commit in the given release milestone.

\def\churn{\mathrm{churn}}

More precisely, let $C$ be a set of commits and $E$ be a set of modules.
First, we define a function $\churn(e, c, c')$ that represents the churn of a module $e \in E$ in a pair of commits $c,c' \in C$.
Given two releases $r$ and $r'$, the sequence of $n$ commits between them is denoted by $c_1, c_2, \dots, c_n$, where $c_r = c_1$ and $c_{r'} = c_n$ are the commits corresponding to the releases $r$ and $r'$, respectively.
\emph{Release-based change size} and \emph{commit-based change size} of module $e$ are denoted by $\Delta^\mathrm{release}_e$ and $\Delta^\mathrm{commit}_e$, respectively, and computed as follows:
\begin{align*}
  \Delta^\mathrm{release}_e &= \churn(e, c_r, c_{r'}), \\
  \Delta^\mathrm{commit}_e  &= \sum_{i = 1}^{n-1} \churn(e, c_i, c_{i+1}).
\end{align*}

\def\LOC{\mathrm{LOC}}

Assuming that the effort to maintain $k$ lines of code is available, we measure these two metrics in the modules within the top-$k$ LOC and then sum the values for each module. 

Let $R=\langle e_1, e_2,\dots\rangle$ be the ranking of modules in the descending order of the change-proneness that the change-prediction model calculated.
Therefore, the number of top results $l$ can be the number that satisfies the following condition:
\begin{align*} 
  \sum_{i=1}^l \LOC(e_i) < k \leq \sum_{i=1}^{l+1} \LOC(e_i)
\end{align*}
where $\LOC(e)$ counts the LOC of module $e$ at release $r$. 
In our hypothetical situation, only the top-$l$ ranking results are utilized because acceptable development effort is limited.

Using the top $l$ items in the ranking $R_l= \langle e_1, e_2,\dots, e_l \rangle$, two change-size metrics can be computed follows:
\begin{align*}
  \text{\textit{top-k release-based change ratio}} &= 
 \frac{\sum_{e \in R_l}\Delta^\mathrm{release}_e}{\sum_{e \in R_l}\LOC(e)}, \\
  \text{\textit{top-k commit-based change ratio}} &= 
 \frac{\sum_{e \in R_l}\Delta^\mathrm{commit}_e}{\sum_{e \in R_l}\LOC(e)}.
\end{align*}
The assumption behind these metrics is that the larger these values are, the more useful the ranking of the top-$k$ lines will be in a realistic context, where the acceptable development effort is limited.
We use 100, 500, 1,000, 5,000, and 10,000 for the value of $k$ for the evaluation to simulate different contexts in terms of the acceptable development effort.

We applied the two-sided Wilcoxon signed-rank test at a significance level of 0.05 To determine whether there were differences in the median values of each evaluation metric.

\subsubsection{Results and Discussion}
\begin{figure}[tb] 
  \centering
  \begin{subfigure}{\linewidth}
  \includegraphics[width=\linewidth]{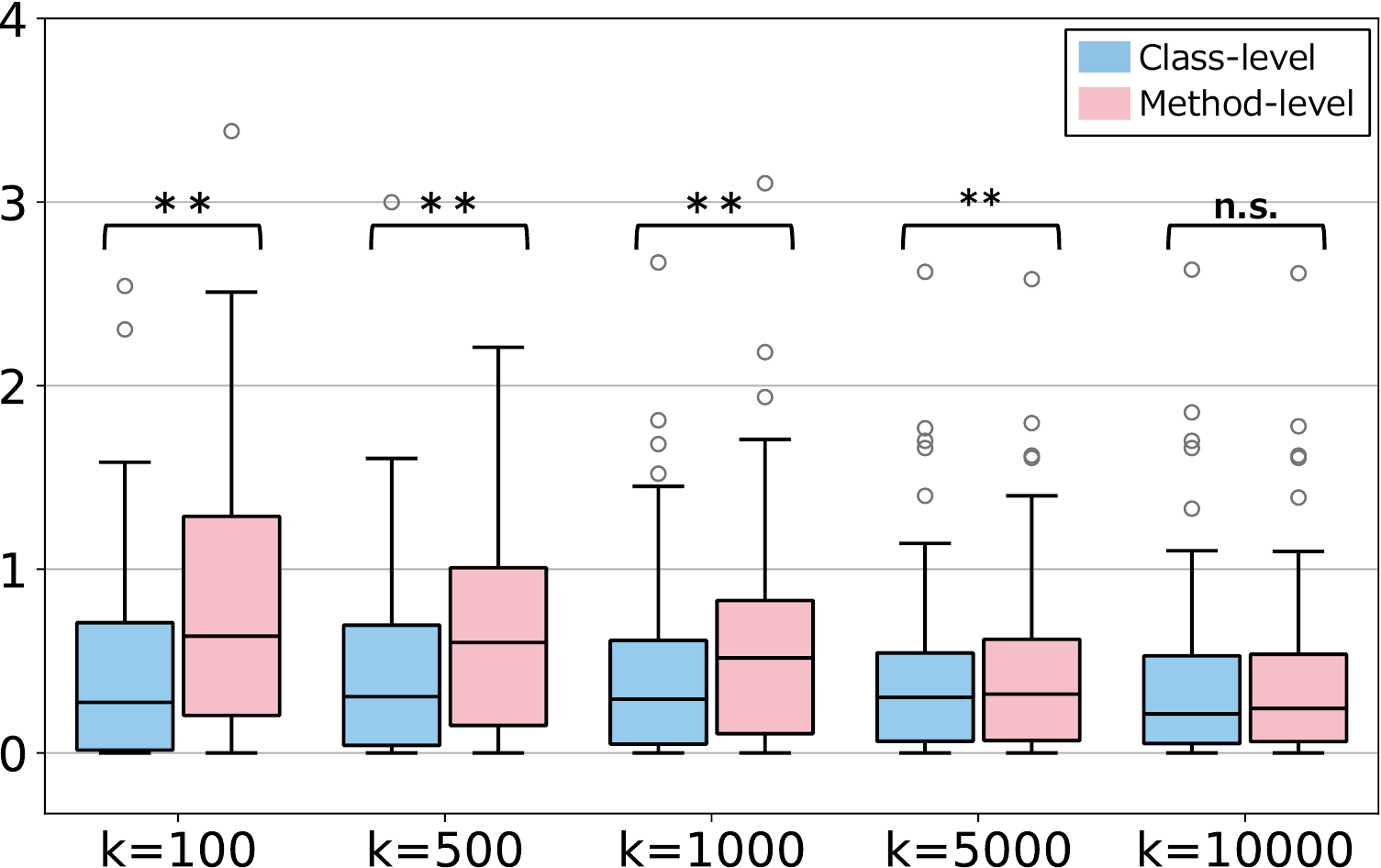} 
  \caption{Top-$k$ release-based change ratio.}\label{fig:diff}
  \end{subfigure} ~ \\

  \begin{subfigure}{\linewidth}
  \centering
  \includegraphics[width=\linewidth]{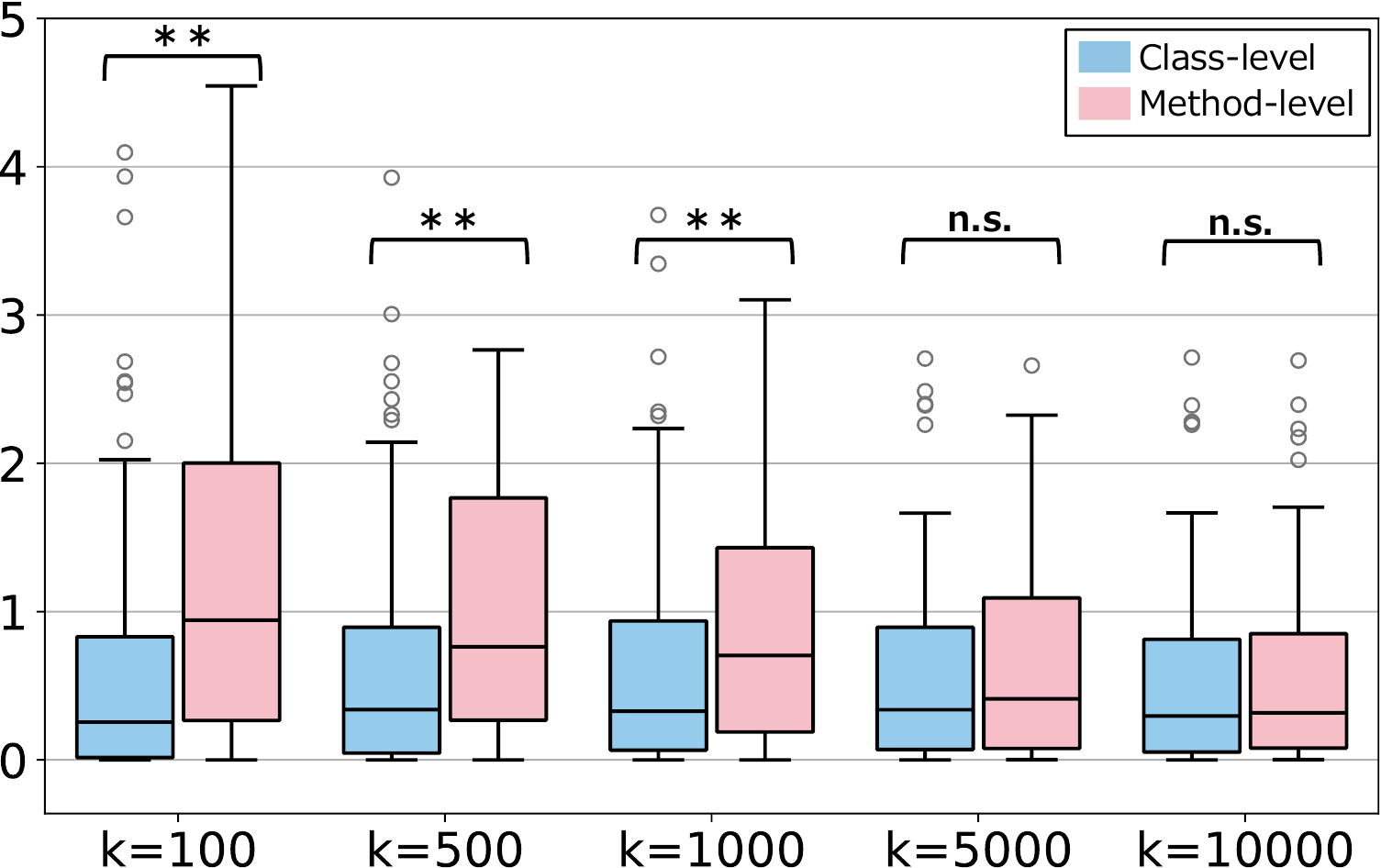} 
  \caption{Top-$k$ commit-based change ratio.}\label{fig:log}
  \end{subfigure}
  \caption{Results for \RQ{3}: effort-aware evaluation.}
  \label{fig:rq3}
\end{figure}

\Cref{fig:rq3} depicts the values of top-$k$ release-based and commit-based change ratios at both method-level and class-level, respectively.
\Cref{tab:diff-results,tab:log-results} present the median values of each metric at each level, the p-values from the Wilcoxon signed-rank test, and the values and effect sizes of Cliff’s delta.

\begin{table}[tb]\centering
    \caption{Median of top-$k$ Release-Based Change Ratio and Test Results}
    \label{tab:diff-results}
    {\tabcolsep=4pt\begin{tabular}{rcccrl} \hline
         $k$ & Class-level & Method-level & p-value & \multicolumn{2}{c}{Cliff's delta} \\ \hline
        100 & 0.19 & 0.66 & $1.1\times10^{-6}$** & 0.37 & (medium) \\
        500 & 0.24 & 0.60 &$1.1\times10^{-6}$**& 0.28 & (small) \\
        1,000 & 0.26 & 0.46 & $1.3\times10^{-5}$** & 0.22 & (small) \\
        5,000 & 0.23 & 0.30 & $0.0019^{}$** & 0.07 & (negligible)\\
        10,000 & 0.20 & 0.23 & 0.45 n.s. & 0.032 & (negligible)\\
        \hline
    \end{tabular}}
\end{table}

\begin{table}[tb]\centering
    \caption{Median of top-$k$ Commit-Based Change Ratio and Test Results} \label{tab:log-results}
    {\tabcolsep=4pt\begin{tabular}{rcccrl} \hline
         $k$ & Class-level & Method-level & p-value & \multicolumn{2}{c}{Cliff's delta} \\ \hline
        100 & 0.25 & 0.94 & $7.7\times10^{-5}$** & 0.34 & (medium) \\
        500 & 0.33 & 0.76 & $3.0\times10^{-4}$** & 0.24 & (small) \\
        1,000 & 0.32 & 0.70 & $2.5\times10^{-5}$** & 0.21 & (small) \\
        5,000 & 0.33 & 0.41 & 0.058 n.s.& 0.064 & (negligible)\\
        10,000 & 0.29 & 0.31 & 0.62 n.s.& 0.032 & (negligible)\\
        \hline
    \end{tabular}}
\end{table}

In both change sizes, smaller values of $k$ were associated with higher performance in method-level prediction.
As $k$ increased, the discrepancy between method-level and class-level predictions decreased.
The most significant difference was observed for $k=100$.
Method-level prediction performed significantly better with medium effect size in top-$k$ release-based change ratio and small in top-$k$ commit-based change ratio.
At $k$ = 10,000, neither top-$k$ release-based nor commit-based change ratios showed any significant differences, and its effect size was negligible.
A possible interpretation for these results is that, as $k$ increases, the modules included within $k$ lines encompass lower-ranking positions, thereby diluting the association with predictive performance.
These results indicate that method-level predictions become more effective when development resources are scarce.

\Conclusion{%
The smaller the value of $k$, the larger the difference between the method-level prediction and class-level prediction.
For $k=100$, method-level prediction significantly outperformed class-level prediction in both top-$k$ release-based and commit-based change ratios, resulting in a median difference of 0.47 and 0.69, respectively.
When development resources are limited, the efficiency of maintenance activities can be enhanced by utilizing method-level prediction.
}

\section{Discussion}\label{s:discussion}
Following the outcomes of \RQ{1} to \RQ{3}, we discuss the effectiveness of method-level change prediction compared with class-level change prediction.
Method-level change prediction is considered effective because it can predict the specific code snippets within the source code where changes occur more directly compared to class-level prediction.
However, there was a concern that the performance of predictions might decrease as the prediction target shifts from the entire project’s classes to methods.
According to the results of \RQ{1}, as anticipated, method-level prediction performance was lower than 
that of class-level prediction.

However, the results of \RQ{2} suggest that method-level prediction may reduce the detection of false positive methods compared with adapting class-level predictions to corresponding methods.
This implies that when developers utilize change prediction results for method-level maintenance, it might be beneficial to focus maintenance efforts on areas where changes are most anticipated while accepting some oversights.

In \RQ{3}, we considered a realistic setting in which maintenance effort was limited.
Our findings suggest that method-level change prediction enables more efficient maintenance when development resources are scarce.

Based on these findings, we conclude that method-level change prediction is more effective than class-level change prediction for prioritizing efficient use of limited maintenance resources rather than aiming for comprehensive maintenance.
Method-level prediction can be used effectively when the developers want to maintain concentrated change locations while allowing some tolerance for missed change locations, or when development resources are limited and large-scale maintenance is not feasible.

\section{Threats to Validity}\label{s:validity}

\subsection{Internal Validity}
In this paper, we used CK\cite{aniche-ck} to collect product metrics and implemented a tool to collect process metrics.
Although CK has been widely used in other change prediction studies and there is no standard tool for calculating process metrics to the best of our knowledge, the results may vary depending on the implementation.

We used a method repository to collect the change history of methods.
This approach enabled the acquisition of method-level change histories using Git.
Farah et al. used ChangeDistiller for the same purpose.
The code differences generated by ChangeDistiller ignore changes to whitespace and comments, whereas those generated by Git do not. 
Moreover, there are differences in the approach to detect renaming.
In addition, ChangeDistiller detects renaming by comparing Abstract Syntax Tree (AST) nodes between different versions of the code, identifying instances where nodes with similar structures but different names indicate a rename\cite{ChangeDistiller}.
In contrast, Historage detects renaming based on the similarity of file contents when no matching file name is found\cite{Historage}.
It compares the content and names of files across versions, detecting renaming through similarity between the original and modified files.
Consequently, there are slight discrepancies in the changes considered by each approach.
However, these facts do not detract from the comparison results between class-level prediction and method-level prediction, which is the focus of this paper.

\subsection{Construct Validity}
The types of independent variables used for change prediction are considered threats to construct validity.
In this paper, product and process metrics were used as independent variables because the approach that employs these two types of metrics has shown the best performance in the paper by Farah et al. and is widely used in other studies\cite{malhotra-survey}.
However, other independent variables have been explored in class-level prediction\cite{catolino-smell,catolino-dev}.
If we introduce these metrics to the independent variables, it might lead to different outcomes.

\subsection{External Validity}
We did not use multiple machine learning models, sets of metrics, or data resampling techniques; we selected the most effective ones, as identified by Farah et al.
Employing different approaches from those presented in this study could yield different results.
However, the primary goal of this paper was to compare class-level predictions with method-level predictions.
These facts do not undermine our discussion from the perspective of comparisons between different prediction levels.

The projects targeted in this paper are all open-source software developed in Java.
This poses a threat to the generalizability of the results of this study to other software programs.

\section{Conclusion and Future Work}\label{s:conclusion}
In this paper, we conducted change prediction experiments at both class-level and method-level using existing change prediction techniques and compared the results.
Regarding \RQ{1}, we found that method-level change prediction underperformed compared to class-level prediction.
For \RQ{2}, we evaluated class-level change prediction at method-level and compared the results with that obtained from method-level change prediction.
The findings indicated that, although method-level predictions resulted in a higher number of false negatives, they reduced the number of false positives.
To address \RQ{3}, we introduced an effort-aware evaluation.
The results showed that method-level prediction is more effective when maintenance resources are limited.

In future research, we are planning to analyze the impact of release size on predictions.
The size of a release affects the proportion of change-prone modules.
This may change the difficulty of making predictions.

All the datasets of the experiments conducted in this paper are publicly available\cite{figshare}.

\section*{Acknowledgments}
This work was partly supported by JSPS Grants-in-Aid for Scientific Research (JP24H00692, JP23K24823, JP21H04877, JP21K18302, and JP21KK0179).

\IEEEtriggeratref{20}
\bibliographystyle{IEEEtran}
\bibliography{references} 

\end{document}